\documentclass[aps,prc]{revtex4}
\usepackage{graphics}

\begin{document}
\title{A Test of X(5) for the $\gamma$ Degree of Freedom}
\author{R. Bijker$^1$, R.F. Casten$^{2,3}$, N.V. Zamfir$^2$,
and E.A. McCutchan$^2$} \affiliation{$^1$Instituto de Ciencias
Nucleares, Universidad Nacional Aut\'onoma de M\'exico,
A.P. 70-543, 04510 M\'exico, D.F., M\'exico \\
$^2$Wright Nuclear Structure Laboratory, Yale
University, New Haven, Connecticut 06520-8124\\
$^3$Insitiut f\"{u}r Kernphysik, Universi\"{a}t zu K\"{o}ln,
K\"{o}ln, GERMANY}

\begin{abstract}
We present the first extensive test of the critical point symmetry
X(5) for the $\gamma$ degree of freedom, based in part on recent
measurements for the $\gamma$-band in $^{152}$Sm. The agreement is
good for some observables including the energies and most intra-
and inter-band transitions, but there is also a serious
discrepancy for one transition.
\end{abstract}
\maketitle

The recent \cite{1} proposal of the critical point symmetry X(5)
describing a vibrator to axial rotor first order phase transition
has introduced a new paradigm into the arsenal of nuclear models
and has generated considerable interest, both experimental and
theoretical. Of course, no nucleus need exhibit exact agreement
with such a symmetry. Since nuclei contain integer numbers of
nucleons, their properties change discretely with $N$ and $Z$, and
a transition region may well by-pass the exact critical point.
Nevertheless, empirical examples of nuclei close to X(5) in
structure were identified in $^{152}$Sm \cite{2} and $^{150}$Nd
\cite{3}. Although X(5) is parameter-free (except for scale), the
overall agreement with the data is quite good. A notable
discrepancy in the absolute scale of inter-band $B(E2)$ values,
discussed in detail in refs.~\cite{2,4}, and very recently, in
\cite{5}, probably reflects the fact that these $N$ = 90 nuclei
are slightly to the rotor side of the phase transition. Recently,
other candidates for X(5) have been discussed \cite{6,7,8,9}.

To date, the X(5) predictions have been discussed primarily for
the yrast and yrare degrees of freedom, that is, for the
quasi-ground band and for the sequence of levels built on the
$0^+_2$ level. However, the solution for the infinite square well
(in $\beta$) ansatz underlying X(5) involves a separation of
variables in the $\beta$ and $\gamma$ degrees of freedom, and
leads to a full set of predictions for the
quasi-$\gamma$-vibrational levels as well.

To date, the most significant comparison of X(5) predictions with
data for the $\gamma$ degree of freedom was presented for
$^{104}$Mo \cite{10}. It includes relative $\gamma$-band energies
and several branching ratios. It is the purpose of the present
Rapid Communication to exploit recent experiments using the GRID
technique at the ILL and polarization measurements of $M1/E2$
mixing ratios at Yale \cite{11} to present an extensive comparison
of X(5) with $\gamma$-band data in $^{152}$Sm, the first nucleus
proposed to exhibit X(5) character. This comparison, including
spins up to $9^+_\gamma$, about 15 absolute $B(E2)$ values, and a
number of independent branching ratios, is the most thorough to
date.

To compare these and other data on the $\gamma$-band in $^{152}$Sm
with X(5) one needs to explicitly solve the X(5) equation in
$\gamma$. The starting point is the Bohr Hamiltonian
\begin{eqnarray}
H = - \frac{\hbar^2}{2B} \left[ \frac{1}{\beta^4}
\frac{\partial}{\partial \beta} \beta^4 \frac{\partial}{\partial
\beta} + \frac{1}{\beta^2 \sin 3\gamma} \frac{\partial}{\partial
\gamma} \sin 3\gamma \frac{\partial}{\partial \gamma}
-\frac{1}{4\beta^2} \sum_{\kappa} \frac{Q_{\kappa}^2(\Omega)}
{\sin^2(\gamma-\frac{2\pi}{3}\kappa)} \right] + V(\beta,\gamma) ~,
\end{eqnarray}
with $V(\beta,\gamma)=V(\beta)+V(\gamma)$ \cite{1}. The potential
in $\beta$ is taken to be an infinite square well with
$V(\beta)=0$ for $\beta \leq \beta_W$ and $V(\beta)=\infty$ for
$\beta > \beta_W$, whereas the potential in $\gamma$ is assumed to
be harmonic around $\gamma_0$ with $V(\gamma)=\frac{1}{2} C
(\gamma-\gamma_0)^2$. Approximate solutions can be obtained in the
limit of small oscillations in the $\gamma$ variable combined with
an adiabatic limit to separate the $\beta$ and $\gamma$ variables.
The energy eigenvalues are given by
\begin{eqnarray}
E(s,n_{\gamma},L,K) = E_0 + a \, (x_{s,\nu})^2 + b \,
(n_{\gamma}+1) ~,
\end{eqnarray}
where $x_{s,\nu}$ is the $s$-th zero of a cylindrical Bessel
function $J_{\nu}(x)$ with
\begin{eqnarray}
\nu=\sqrt{\frac{L(L+1)-K^2}{3}+\frac{9}{4}} ~.
\end{eqnarray}
For $K=0$ this form reduces to the result obtained in \cite{1}. We
note that this solution is valid for both the prolate
($\gamma_0=0$) and the oblate case ($\gamma_0=\pi/3$) due to the
appearance of the irrotational moments of inertia in the Bohr
Hamiltonian which vanish about the symmetry axes. There is an
important difference with the moments of inertia of a rigid rotor,
for which the relative sign of the $L(L+1)$ and $K^2$ terms would
depend on whether the deformation is prolate or oblate \cite{12}.

$B(E2)$ values can be obtained from the matrix elements of the
quadrupole operator
\begin{eqnarray}
T^{E2}=t\beta\left[ {\cal D}^{(2)}_{\mu,0}(\Omega) \cos \gamma +
\frac{1}{\sqrt{2}} \left( {\cal D}^{(2)}_{\mu,2}(\Omega) + {\cal
D}^{(2)}_{\mu,-2}(\Omega) \right) \sin \gamma \right] ~.
\label{TE2}
\end{eqnarray}
The first term describes $\Delta K=0$ transitions and the second
one $\Delta K=2$ transitions. The calculation of matrix elements
of the quadrupole operator involves an integral over the Euler
angles $\Omega$, and over the deformation variables $\beta$ and
$\gamma$
\begin{eqnarray}
B(E2;s n_{\gamma} L K \rightarrow s' n_{\gamma}' L' K')=
\frac{5}{16\pi} \left< L,K,2,K'-K|L',K' \right>^2 \,
B^2_{s\nu;s'\nu'} \, C^2_{n_{\gamma}K;n_{\gamma}'K'} ~,
\label{be2}
\end{eqnarray}
where $B_{s\nu;s'\nu'}$ contains the integral over $\beta$
\cite{1}, and $C_{n_{\gamma}K;n_{\gamma}'K'}$ over $\gamma$. In
the derivation of the $B(E2)$ values of Eq.~(\ref{be2}) we have,
just as for the energies, assumed small oscillations in $\gamma$.
For $\Delta K=0$ transitions the $\gamma$-integral reduces to the
orthonormality condition of the wave functions in $\gamma$, i.e.
$C_{n_{\gamma}K;n_{\gamma}'K}=\delta_{n_{\gamma},n_{\gamma}'}$,
whereas for $\Delta K=2$ transitions this integral can be
interpreted as an intrinsic transition matrix element.

The four independent coefficients that enter in the calculations,
$B$, $\beta_W$, $C$ and $t$ can be determined from two excitation
energies, e.g. $E_{2_1}$ and $E_{2_{\gamma}}$, and two B(E2)
values for $\Delta K=0$ and $\Delta K=2$ transitions, $e.g.$,
$B(E2;2_1 \rightarrow 0_1)$ and $B(E2;2_{\gamma} \rightarrow
0_1)$.

In Table I we present the X(5) results for the energies of the
first two bands with $n_{\gamma}=0$, $K=0$, and the $\gamma$-band
with $n_{\gamma}=1$, $K=2$. The moments of inertia of the ground
band and the $\gamma$-band are almost identical, and much larger
than that of the first excited $K=0$ band. The $B(E2)$ values for
intraband transitions in the $\gamma$-band are shown in Table II,
and those for interband transitions to the ground and 0$^+_2$
bands in Tables III and IV. The values are normalized to the
$\Delta K=0$ ground band transition $B(E2;2_1 \rightarrow
0_1)=100$ and the $\Delta K=2$ transition $B(E2;2_{\gamma}
\rightarrow 0_1)=10$, respectively. Many of these results are
illustrated in Fig.~1. It is interesting to note that while the
relative $\gamma-\gamma$ and $\gamma$-ground band $B(E2)$ values
are close to the Alaga rules, the $\gamma \rightarrow$ 0$^+_2$
band values differ significantly.

Fig. 2 and Table V give a comparison with X(5) for all known
quasi-$\gamma$-band energies and absolute $B(E2)$ values in
$^{152}$Sm. Table VI presents a comparison of the data with X(5)
for cases where relative $B(E2)$ values are known. The data in
these tables and Fig.~2 are taken from
refs.~\cite{11,13,14,15,16}. The comparisons in Tables V, VI and
Fig.~2, like other X(5) predictions, are parameter-free except for
scale. As mentioned before, for the $\gamma$ degree of freedom,
there are two additional scales that must be fixed beyond the
normalization in Ref.~\cite{2} for the yrast and yrare levels.
Thus, in Table V and Fig.~2 (left) we have normalized
$E(2^+_\gamma)$ to the experimental value and the $B(E2)$ values
for the $\Delta K=2$ transitions to an average of the $2^+_\gamma
\rightarrow 0^+_1$ and $2^+_\gamma \rightarrow 2^+_1$ transitions.
The $\gamma \rightarrow \gamma$ in-band transitions have the same
normalization as for the $\Delta K=0$ transitions among the yrast
and yrare levels in ref.~\cite{2}, and are not affected by the
scale factor for the $\Delta K=2$ transitions.

The results are quite interesting. First, they provide an
extensive test of X(5) for the $\gamma$-degree of freedom.
Secondly, they exhibit both excellent agreement and at least one
severe discrepancy. X(5) agrees quite well with the data for the
$\gamma$-band energies, and far better than other paradigms such
as the the axial rotor, as seen in Fig.~3. As with the yrast and
yrare levels, $^{152}$Sm deviates from X(5) slightly in the
direction of the rotor. The spacings $within$ the odd-even spin
couplets ($3^+_\gamma$, $4^+_\gamma$), ($5^+_\gamma$,
$6^+_\gamma$), ($7^+_\gamma$, $8^+_\gamma$) are almost exact while
the spacings $between$ couplets are slightly smaller in X(5)
compared with the data.

Turning to transition rates, the three known intraband $B(E2)$
values in the quasi-$\gamma$-band are reasonably consistent with
the data, and the $B(E2)$ values to the ground band are in rather
good agreement. Of these latter transitions, the agreement is
poorest for the $4^+_\gamma \rightarrow 2^+_1$, transition [0.59
(17) W.u. experimentally compared to 2.86 W.u.~in X(5)]. Most of
the transitions to the yrare, or $0^+_2$-band, levels are,
experimentally, very weak (or else only upper limits are known),
and so are the X(5) predictions. However, there is one glaring
discrepancy, namely for the $2^+_\gamma \rightarrow 2^+_2$
transition, whose measured $B(E2)$ value \cite{16} is 27(4) W.u.,
while X(5) predicts 0.20 W.u. The origin of this problem may be
that, in the X(5) solution, the $\beta$ and $\gamma$ degrees of
freedom are separated. In fact, calculations with both the IBA
\cite{16} and GCM \cite{17} models, where their coupling is
included, predict much higher $B(E2;2^+_\gamma \rightarrow 2^+_1)$
values, which actually exceed the experimental ones. We noted
earlier that the $\gamma \rightarrow$ 0$^+_2$ band $B(E2)$ values
differ significantly from the Alaga rules. The branching ratio
from the 4$^+_\gamma$ level ($<$ 0.7) is consistent with X(5)
(0.84) but differs from the Alaga rule (2.93). It would clearly be
of interest to measure relative $B(E2)$ values from higher lying
members of the quasi-$\gamma$-band to further test the X(5)
predictions.

Finally, the comparison of branching ratios in Table VI (where
absolute rates are not known or poorly known) shows mixed
agreement. The very small values, which are ratios of interband
transitions to the ground band to intra-quasi-$\gamma$ band
transitions, are likewise very small in X(5) and in good agreement
with the data. However, for the three cases of branching ratios to
the ground band, the experimental ratios are about 3--6 times
larger than in X(5).

Overall, considering that X(5) is an invariant paradigm based on
an infinite square well potential in $\beta$ and a harmonic
potential in $\gamma$ that is parameter free (except for scale),
the agreement is quite good. At the same time, the striking
disagreement for the $2^+_\gamma \rightarrow 2^+_2$ transition
needs to be better understood. Another area worth investigating
are other forms of $V(\beta)$ \cite{18} and/or $V(\gamma)$, in
particular their effects on energies and $B(E2)$ values.

\section*{Acknowledgments}
We would like to thank F.~Iachello and N.~Pietralla for useful
discussions. This work is supported in part by a grant from
CONACyT, M\'exico, and by USDOE Grant No. DE-FG02-91ER-40609.

\clearpage

\begin{table}
\begin{tabular}{|cccc|}\hline
$L$  &$E_1$($n_\gamma$ = 0) &$E_2$($n_\gamma$ = 0)  &$E_1$($n_\gamma$ = 1) \\
\hline
0   &0  &565 &\\
2   &100    &745    &1000 \\
3   &   &    &1094\\
4   &290   &1069    &1204\\
5   &   &    &1327\\
6   &543   &1475    &1464\\
7   &   &    &1613\\
8   &848   &1944    &1774\\
9   &   &    &1946\\
10  &1203   &2469    &2131\\
11  &   &    &2327\\
12  &1604   &3045    &2534\\
13  &   &    &2753\\
14  &2051   &3672    &2983\\
15  &   &    &3225\\
16  &2544   &4348    &3477\\
\hline
\end{tabular}
\caption{Excitation energies in X(5). The energies are normalized
to E$_{2_1}$ = 100 and E$_{2_\gamma}$ = 1000.}
\end{table}

\begin{table}
\begin{tabular}{|ccc|}\hline
$L_\gamma \rightarrow$  &($L$ - 2)$_\gamma$ &($L$ - 1)$_\gamma$ \\
\hline
3$_\gamma$    &   &186 \\
4$_\gamma$    &63 &151 \\
5$_\gamma$    &110   &116 \\
6$_\gamma$    &144   &91 \\
7$_\gamma$    &172   &72 \\
8$_\gamma$    &194   &59 \\
9$_\gamma$    &212   &49 \\
10$_\gamma$   &227   &42 \\
\hline
\end{tabular}
\caption{B(E2) values for transitions within the
quasi-$\gamma$-band for X(5). These values are normalized to the
ground band transition B(E2; 2$_1$ $\rightarrow$ 0$_1$) = 100.}
\end{table}

\begin{table}
\begin{tabular}{|cccccc|}\hline
$L_\gamma \rightarrow$  & ($L$ - 2)$_1$ & ($L$ - 1)$_1$ & $L_1$ &
($L$ + 1)$_1$   & ($L$ + 2)$_1$ \\ \hline
2$^+_\gamma$    &10  &   &15   &   &0.78 \\
3$^+_\gamma$    &   &20   &   &8.4  & \\
4$^+_\gamma$    &6.8  &   &22   &   &1.9 \\
5$^+_\gamma$    &   &20   &  &12   &  \\
6$^+_\gamma$    &6.3  &   &25   &   &2.7 \\
7$^+_\gamma$    &   &21   &   &14   &  \\
8$^+_\gamma$    &6.2  &   &27   &   &3.3 \\
9$^+_\gamma$    &   &22   &  &16   &  \\
10$^+_\gamma$   &6.3  &   &29   &   &3.7 \\
\hline
\end{tabular}
\caption{B(E2) values for transitions from the quasi-$\gamma$-band
to the ground band for X(5). These values are normalized to the
transition B(E2; 2$_\gamma$ $\rightarrow$ 0$_1$) = 10.}
\end{table}

\begin{table}
\begin{tabular}{|cccccc|}\hline
$L_\gamma \rightarrow$  & ($L$ - 2)$_2$ & ($L$ - 1)$_2$ & $L_2$ &
($L$ + 1)$_2$   & ($L$ + 2)$_2$ \\ \hline
2$^+_\gamma$    & 0.89  &   &0.48   &   &0.001 \\
3$^+_\gamma$    &   &1.29   &   &0.081  & \\
4$^+_\gamma$    & 0.73  &   &0.61   &   & 0.002 \\
5$^+_\gamma$    &   &1.10   &  &0.10   &  \\
6$^+_\gamma$    & 0.55  &   &0.58   &   & 0.003 \\
7$^+_\gamma$    &   &0.91   &   &0.11   &  \\
8$^+_\gamma$    & 0.42  &   &0.52   &   & 0.004 \\
9$^+_\gamma$    &   &0.75   &  &0.11   &  \\
10$^+_\gamma$   & 0.33  &   &0.46   &   & 0.005 \\
\hline
\end{tabular}
\caption{B(E2) values for transitions from the quasi-$\gamma$-band
to the 0$^+_2$ band for X(5). These values are normalized to the
transition B(E2; 2$_\gamma$ $\rightarrow$ 0$_1$) = 10.}
\end{table}

\begin{table}
\begin{tabular}{|r|c|c|}\hline
{$J_i \rightarrow J_f$}   & \multicolumn{2}{c|}{$B(E2)$~W.u.} \\
\cline{2-3}
   &Exp.  &X(5) \\ \hline
2$^+_\gamma \rightarrow$ 0$^+_1$    & 3.62 (17) & 4.20 \\
$2^+_1$ & 9.3 (5) &   6.33 \\
4$^+_1$ & 0.78 (5)  & 0.33 \\
0$^+_2$ & $<$ 0.05  & 0.37 \\
2$^+_2$ & 27 (4)    & 0.20 \\ \hline 3$^+_\gamma \rightarrow$
2$^+_1$ & 7 - 17    & 8.31 \\
4$^+_1$ & 7 - 18    & 3.51 \\
2$^+_2$ & $<$ 0.52  & 0.54 \\
2$^+_\gamma$ & 62 - 798    & 267.2 \\ \hline
4$^+_\gamma \rightarrow$ 2$^+_1$    & 0.59 (17) &   2.85   \\
4$^+_1$ & 5.5 (16)  & 9.05 \\
6$^+_1$ & 1.2 (4)   & 0.80 \\
2$^+_2$ & 0.18 (7)  & 0.31 \\
4$^+_2$ & $<$ 35    & 0.26 \\
2$^+_\gamma$    & 50 (15)  & 90.7 \\
3$^+_\gamma$ & $<$ 250  & 217.3 \\ \hline
\end{tabular}
\caption{Comparison of absolute $B(E2)$ values for the
quasi-$\gamma$-band in $^{152}$Sm with X(5). The scale for $\Delta
K=2$ transitions is normalized to approximately reproduce the
$2^+_\gamma \rightarrow 0^+_1$ and $2^+_\gamma \rightarrow 2^+_1$
B(E2) values. Data are from refs.~\cite{5,7,8,9,10}.}
\end{table}

\begin{table}
\begin{tabular}{|c|c|c|}\hline
$B(E2)$ Ratio   &   Exp.    &X(5)   \\  \hline 3$^+_1\rightarrow$
4$^+_1$ / 3$^+_1 \rightarrow$ 2$^+_1$ & 1.08
(1) & 0.42 \\
5$^+_\gamma \rightarrow$ 4$^+_1$ / 5$^+_\gamma \rightarrow$
3$^+_\gamma$ & 0.039 (13) &   0.054   \\
6$^+_\gamma \rightarrow$ 6$^+_1$ / 6$^+_\gamma \rightarrow$ 4$^+_1$ & 23 (8) & 3.95  \\
7$^+_\gamma \rightarrow$ 8$^+_1$ / 7$^+_\gamma \rightarrow$
6$^+_1$ & 4.1 (14) &   0.69    \\
7$^+_\gamma \rightarrow$ 6$^+_1$ / 7$^+_\gamma \rightarrow$
5$^+_\gamma$ & 0.0099 (17) & 0.036   \\
9$^+_\gamma \rightarrow$ 10$^+_1$ /
9$^+_\gamma \rightarrow$ 7$^+_\gamma$ & 0.021 (21) &   0.022   \\
\hline
\end{tabular}
\caption{Comparison of B(E2) branching ratios (where $M1/E2$
mixing ratios are known) for the quasi-$\gamma$-band in $^{152}$Sm
with X(5) (for levels with unknown (or poorly known) lifetimes or
absolute $B(E2)$ values). Data are from refs.~\cite{5,7,8,9}.}
\end{table}
\clearpage

\begin{figure}
\centering
\includegraphics{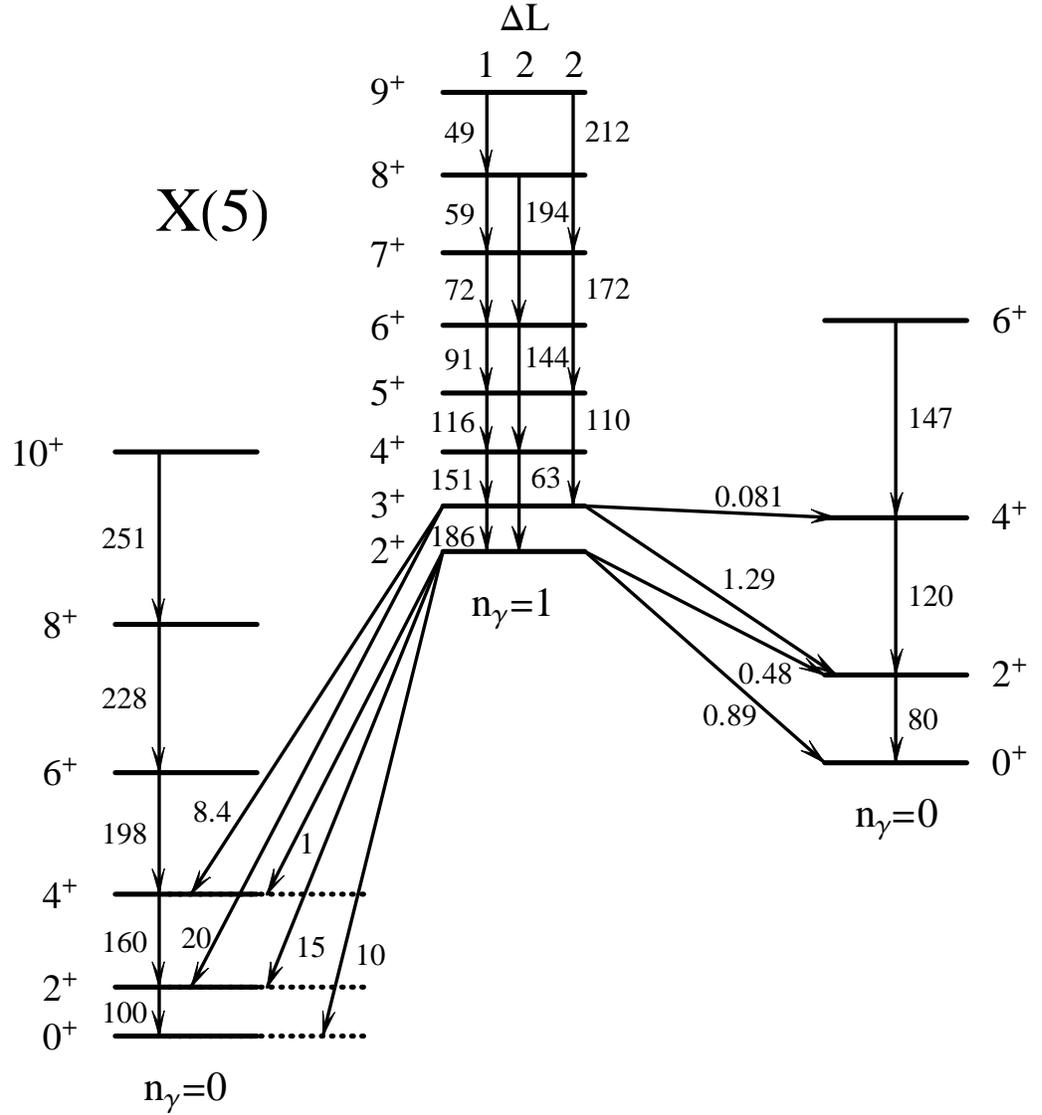}
\vspace{-7cm} 
\caption[]{Predictions of X(5). The energy and $B(E2)$
values are normalized as in Tables I - IV. Here and in Fig.~2, the
numbers on the transition arrows are $B(E2)$ values.}
\end{figure}

\begin{figure}
\centering 
\includegraphics{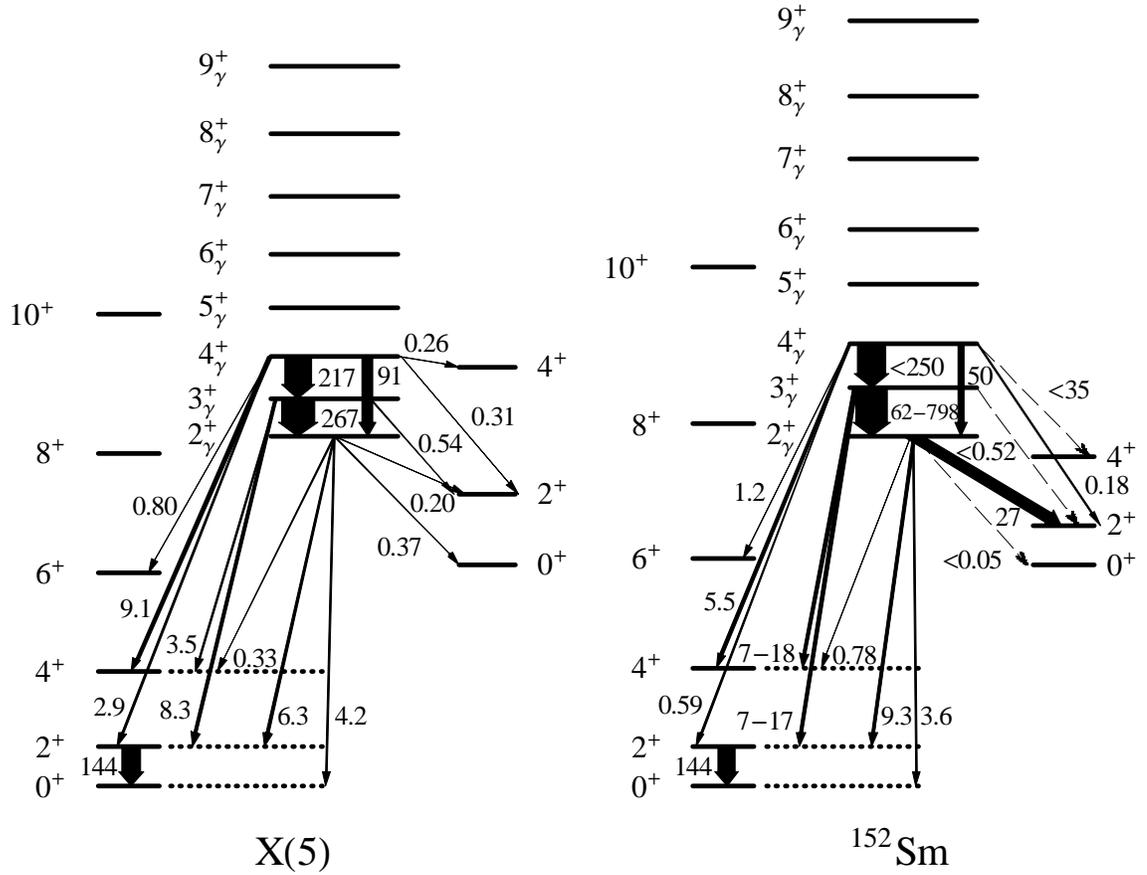}
\vspace{-11cm}
\caption[]{Comparison of the data for the
quasi-$\gamma$-band in $^{152}$Sm with X(5) predictions. Data from
refs.~\protect\cite{11,13,14,15,16}. See text. The thickness of the transition
arrows indicates the corresponding $B(E2)$ values except that, for
readability, the intraband transitions are scaled down in
thickness by a factor of three. The numbers on the transition
arrows are $B(E2)$ values in W.u. The dashed arrows on the right
indicate transitions where only upper limits on the $B(E2)$ values
are known.}
\end{figure}

\begin{figure}
\vspace{3cm}
\hspace{-2cm}\includegraphics{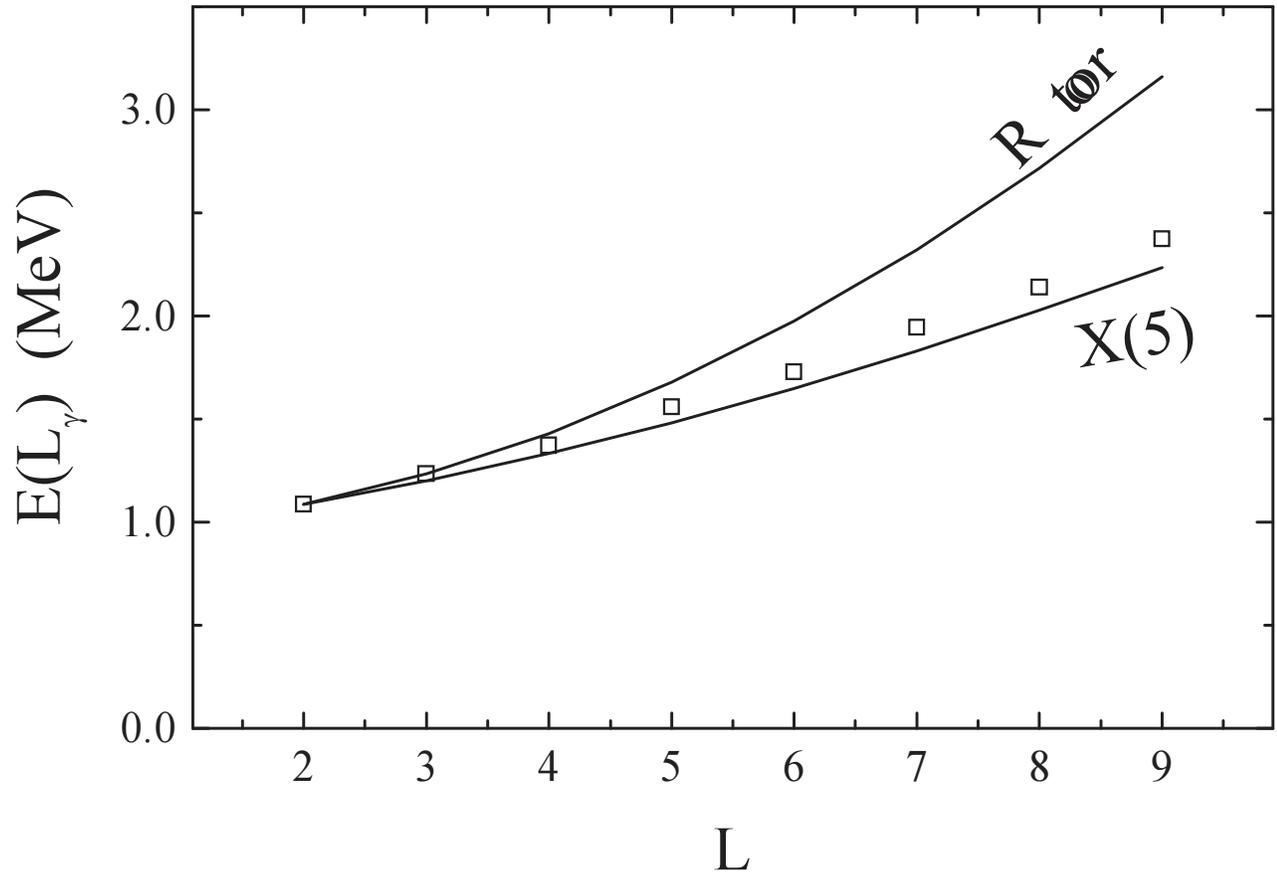} 
\vspace{-9cm}
\caption[]{Comparison of relative $\gamma$-band
energies in $^{152}$Sm with X(5) and with an axial rotor.}
\end{figure}
\end{document}